%% file: main.tex
\documentclass[a4paper,11pt]{article}
\pdfoutput=1 % if your are submitting a pdflatex (i.e. if you have
\usepackage{jcappub} % for details on the use of the package, please
\usepackage[T1]{fontenc} % if needed
\usepackage{gensymb}

\usepackage{accents}
\newlength{\dhatheight}
\newcommand{\doublehat}[1]{
    \settoheight{\dhatheight}{\ensuremath{\hat{#1}}}
    \addtolength{\dhatheight}{-0.25ex}
    \hat{\vphantom{\rule{1pt}{\dhatheight}}
    \smash{\hat{#1}}}}

\title{Dark matter indirect detection limits from complete annihilation patterns}

\author[a]{C.\ Armand}
\author[b]{~and~~B.\ Herrmann}

\affiliation[a]{Department of Astronomy, University of Geneva, Chemin Pegasi 51, CH-1290 Versoix, Switzerland}
\affiliation[b]{LAPTh, Univ.\ Savoie Mont Blanc, CNRS, F-74000 Annecy, France}

\emailAdd{celine.armand@unige.ch}
\emailAdd{herrmann@lapth.cnrs.fr}

\abstract{While cosmological and astrophysical probes suggest that dark matter would make up for 85\% of the total matter content of the Universe, the determination of its nature remains one of the greatest challenges of fundamental physics. Assuming the $\Lambda$CDM cosmological model, Weakly Interacting Massive Particles would annihilate into Standard Model particles, yielding $\gamma$-rays, which could be detected by ground-based telescopes. Dwarf spheroidal galaxies represent promising targets for such indirect searches as they are assumed to be highly dark matter dominated with the absence of astrophysical sources nearby. Previous studies have led to upper limits on the annihilation cross-section assuming single exclusive annihilation channels. In this work, we consider a more realistic situation and take into account the complete annihilation pattern within a given particle physics model. This allows us to study the impact on the derived upper limits on the dark matter annihilation cross-section from a full annihilation pattern compared to the case of a single annihilation channel. We use mock data for the Cherenkov Telescope Array simulating the observations of the promising dwarf spheroidal galaxy Sculptor. We show the impact of considering the full annihilation pattern within a simple framework where the Standard Model of particle physics is extended by a singlet scalar. Such a model shows new features in the shape of the predicted upper limit which reaches a value of $\langle \sigma v \rangle = 3.8\times10^{-24}~\rm{cm}^{-3}\rm{s}^{-1}$ for a dark matter mass of 1~TeV at 95\% confidence level. We suggest considering the complete particle physics information in order to derive more realistic limits.}

\begin{document}
\maketitle
\flushbottom

% ===================================================================
\input{intro}
\input{indirect_detection}
\input{astro_model}
\input{statistical_analysis}
\input{particle_model}

\input{results}
\input{conclusion}

% ===================================================================
\acknowledgments
The authors thank V. Poireau for his useful advice on the analysis, G.\ B\'elanger and F.\ Boudjema for useful discussions, as well as F.\ Calore, Y.\ G\'enolini, and P.\ Salati for their insightful suggestions and comments on the manuscript. This work is supported by {\it Fonds National Suisse} and {\it Investissements d’avenir}, Labex ENIGMASS, contrat ANR-11-LABX-0012. Our numerical analysis coded in {\tt Python} and {\tt C/C++} has been performed on the interactive servers of the Univ.\ Savoie Mont Blanc -- CNRS/IN2P3 MUST computing center and the Yggdrasil servers of the University of Geneva. The figures in this paper have been produced using {\tt Matplotlib} \cite{MatPlotLib}.

% ===================================================================
\appendix
\input{App_spectra_cirelli}
\input{App_CTA_comparison}

% ===================================================================
\bibliographystyle{JHEP}
\bibliography{references}

\end{document}

%% file: intro.tex
% ===================================================================
\section{Introduction}

Numerous observational probes indicate that about 85\% of the total matter of the Universe is composed of non-baryonic cold dark matter (DM). This exotic form of matter is responsible for many phenomema at different scales such as the formation of the large structures, the motion of galaxies and clusters, and the bending of the path of light. In addition to astrophysical evidence, the presence of dark matter is confirmed by cosmological measurements. More precisely, within the cosmological $\Lambda$CDM model, the relic density of cold dark matter (CDM) has been restricted to the rather narrow interval
\begin{equation}
    \Omega_{\rm CDM}h^2 ~=~ 0.1200 \pm 0.0012
    \label{Eq:omh2Planck}
\end{equation}
by combining {\it Planck} data with additional cosmological observations \cite{Planck:2018vyg}. However, the exact nature of dark matter still remains a mystery and represents one of the leading questions in modern particle and astroparticle physics. A popular assumption is that cold dark matter consists of so-called Weakly Interacting Massive Particles (WIMPs) that are predicted by many extensions of the Standard Model (SM). Such a particle is supposed to be stable, massive, and interacts only through weak and gravitational interactions.

Experimentally, the nature of WIMP dark matter can be challenged by different approaches: production at colliders, direct detection, or indirect detection. In the present work, we focus on the latter and assume that WIMPs annihilate into SM particles (bosons, quarks, leptons), which in turn hadronise and/or decay into stable particles such as $\gamma$ rays. The corresponding signals might be detected by $\gamma$-ray telescopes and can be used as probes for indirect DM searches \cite{Feng:2010gw, Hooper:2009zm, MultiMessengers}. High-energy $\gamma$ rays present several advantages compared to charged particles as they do not get deflected by the Galactic magnetic field, and hence their source of emission can be well localised in the sky. Moreover, $\gamma$ rays do not experience significant energy losses during their propagation at Galactic scales. These properties allow us to point directly our $\gamma$-ray instruments to the sources in order to search for signals reaching the Earth.

A vast choice of targets is available for DM indirect searches. We look at DM-rich environments where the DM annihilation rate is the highest to maximise the chance of detection of a possible DM signal. The selection of ideal targets requires a balance between a high enough $J$-factor and dealing with the potential astrophysical $\gamma$-ray background.

One of the most promising targets for DM annihilations are the dwarf spheroidal galaxies (dSphs), satellites of the Milky Way Galaxy. These sources lie at ${\cal O}$(100 kpc) galactocentric distance at high latitudes and hence away from the Galactic plane. They are host to a small amount of luminous mass made of old stellar populations and do not contain much gas or dust. Therefore, new star formation is impossible and dSphs are left with an old stellar population of red giants only. Moreover, dSphs are non-rotating objects but rather are pressure-supported as their kinematics are dominated by the random motion of the stars whose amplitudes are driven by the gravitational potential of the galaxy \cite{Biney_book}. The measurements of the galactic dynamics are based on the line-of-sight velocity of individual stars from which a velocity dispersion profile is derived \cite{Walker:2007ju} to constrain the dark matter distribution profile. Their high mass and low luminosity indicate that the dSphs are DM-dominated with negligible astrophysical background \cite{Cowan}.

Numerous studies based on data from dSphs obtained from several $\gamma$-ray telescopes have been performed in order to identify a potential excess stemming from DM annihilation. They cover different energy ranges starting from a few tens of MeV with Fermi-LAT \cite{Fermi-LAT:2009ihh} up to several tens of TeV with the Air Cherenkov telescopes such as H.E.S.S.\ \cite{HESS:2006fka}, MAGIC \cite{MAGIC:2014zas}, or VERITAS \cite{Park:2015ysa} and the water Cherenkov detector HAWC \cite{Abeysekara:2017mjj}.

In the absence of any excess in the data over the estimated $\gamma$-ray background, only upper limits on the dark matter annihilation cross-section have been derived as a function of the presumed DM mass. These limits are obtained from either one dSph or by performing a stack of their respective observations with either a continuous spectrum \cite{Fermi-LAT:2015att, Fermi-LAT:2016uux, HESS:2014zqa, HESS:2010epq, HAWC:2017mfa, Viana:2011qaf, HESS:2007ora, Aleksic:2013xea, MAGIC:2017avy, MAGIC:2021mog, VERITAS:2017tif, HESS:2020zwn, HESS:2021zzm} or a mono-energetic line \cite{Fermi-LAT:2015kyq, HESS:2018kom, HESS:2020zwn, HESS:2021zzm}. More recently, combined DM searches have been carried out to increase the statistics and the sensitivity to a potential DM signal. Their results present more constraining upper limits than those of individual experiments \cite{MAGIC:2016xys, Alvarez:2020cmw, Hess:2021cdp}. 

The current astrophysical constraints on DM annihilation cross-section are mainly derived assuming one single annihilation channel, e.g.\ annihilation into $W$-bosons or $\tau$-leptons. The main goal of this work is to explore to which extent such limits are affected when relaxing this assumption, i.e.\ taking into account the full annihilation pattern of the presumed DM particles. Contrary to previous studies, we intend to quantify the impact of this more precise procedure on the obtained limits and point out the importance of taking into account the full underlying particle physics model. A secondary goal is to compare our obtained upper limits to the annihilation cross-section predicted in the respective particle physics model. 
In a similar context, a recent study of the Cherenkov Telescope Array (CTA) sensitivities to two classes of dark matter portal models has been published in Ref.\ \cite{Duangchan:2022jqn}.

We simulate mock data for CTA and perform a statistical analysis to derive constraints on the DM annihilation cross-section within the singlet scalar dark matter model as an example of a complete particle physics framework. In particular, we compare the results to those obtained assuming only the individual annihilation channels. We focus on the dwarf spheroidal galaxy Sculptor, which is selected for the DM search programme of CTA \cite{CTAConsortium:2017dvg}. 
 
This article is organised as follows: in Sec.\ \ref{sec:exp_flux}, we recall the $\gamma$-ray flux computation and its key components. In Sec.\ \ref{sec:Astro_model}, we describe the properties of the considered dwarf spheroidal galaxy Sculptor as well as the CTA mock data simulations. We explain our statistical analysis in Sec.\ \ref{sec:stat_analysis}. Then, Sec.\ \ref{Sec:SingletScalarModel} presents the particle physics model that we use in this study. We discuss the obtained results in Sec.\ \ref{sec:results} and conclude on this work in Sec.\ \ref{sec:conlusions}.

%% file: indirect_detection.tex
% ---------------------------------------------------------------------
\section{Expected $\gamma$-ray flux} \label{sec:exp_flux}

In the framework of dark matter (DM) indirect detection, WIMPs annihilate into Standard Model particles which subsequently hadronise and/or decay into observable particles such as $\gamma$ rays. Particular interest is generally given to hadronisation into neutral pions, which decay almost exclusively into photons. The expected differential $\gamma$-ray flux generated by DM annihilation is given by \cite{Bertone:2004pz, armand:tel-03560610}
\begin{equation}
    \frac{{\rm d}\Phi_{\gamma}}{{\rm d}E_{\gamma}} ~=~ \frac{1}{\xi} \frac{\langle\sigma v\rangle}{4\pi m_{\chi}^2} \, \sum_f B_f \frac{{\rm d}N_{\gamma}^f}{{\rm d}E_{\gamma}} \times J \,,
    \label{Eq:dPhidE}
\end{equation}
where the sum runs over all possible annihilation channels. The prefactor $1/\xi$ depends on the nature of the DM particle: $\xi = 2$ if the DM is its own anti-particle (e.g.\ Majorana fermion or neutral scalar), $\xi = 4$ otherwise (e.g.\ Dirac fermion).

We distinguish two key components: the particle physics factor (before the multiplication sign) carries the DM annihilation cross-section averaged over the velocity distribution $\langle\sigma v\rangle$, the DM particle mass $m_{\chi}$, and the differential spectrum ${\rm d}N_{\gamma}^f/{\rm d}E_{\gamma}$ of each annihilation channel $f$ weighted by their respective branching ratio $B_f$. More details on the particle physics model that we use in this study will be given in Sec.\ \ref{Sec:SingletScalarModel}.

The second term (after the multiplication sign) is the so-called astrophysical $J$-factor which describes the DM distribution and the amount of DM annihilations within the source, i.e.\ it quantifies the strength of the signal emitted by the DM annihilations. The $J$-factor is defined as
\begin{equation}
  J ~=~ \iint \: \rho_{\rm{DM}}^2\big(r(s, d, \theta)\big) \:{\rm d}s \: {\rm d}\Omega \,,   
  \label{dwarf_J} \\
\end{equation}
where $\rho_{\text{DM}}$ is the DM density distribution profile defined as a function of the distance $r$ between the centre of the source and the observer. Here, $\Omega$ is the solid angle associated to the source. The distance $r$ can also be expressed as $r^2(s, d, \theta) = s^2+d^2-2sd\cos\theta$, where $s$ is the distance from Earth along the light of sight and $\theta$ is the angular distance with respect to the centre of the source. The quantity $d$ is the distance between the Earth and the nominal position of the source. We note that the derivation of the density distribution profile $\rho_{\textrm{DM}}$ is performed through the Jeans analysis using the spherical Jeans equation formalism \cite{Bonnivard:2014kza, binney2011galactic, Bonnivard:2015xpq}. This method makes use of the spectroscopic data to reconstruct galactic dynamics under the assumptions that the dSphs under consideration are in steady-state hydrodynamic equilibrium, have a spherical symmetry, and are non-rotating objects.

%% file: astro_model.tex
% ===================================================================
\section{Simulated observations of Sculptor with the Cherenkov Telescope Array} 
\label{sec:Astro_model}

To study the impact of a more complex particle model on the resulting upper limits, we produce mock data that simulate the observations expected for the dSph Sculptor with the Cherenkov Telescope Array (CTA). CTA is a telescope array currently under construction. It is divided in two sites, one in La Palma on the Canary Islands in the Northern hemisphere and the second in the Atacama desert in Chile located in the Southern hemisphere. The array will cover an energy range between 20~GeV and 300~TeV and will consist of a total of about a hundred telescopes of three different sizes including large sized telescopes to capture the lowest energy $\gamma$ rays, medium-sized telescopes to cover the core energy range, and small-sized telescopes covering the highest energy events. The average Point Spread Function (PSF) of the instrument is designed to be about $0.1\degree$. CTA will observe the most promising dSphs with the highest $J$-factor as part of its upcoming dark matter search programme starting with Sculptor and Draco, one in each hemisphere.

We make use of the \texttt{Gammapy} distribution package \cite{gammapy:website} for our mock data production. We focus on the case of upper limit derivation, where no signal from dark matter is detected. We build our model with no significant excess towards the source of interest and simulate the resulting mock data from \textit{wooble mode} observations of 500 hours total. The wooble mode corresponds to an observation strategy where the telescopes point in a direction offset by a small angle, typically $0.5\degree$, from the nominal source position. The source is observed using four pointing positions alternating the offset in the positive and negative declination and right ascension. This method allows a simultaneous estimate of the background thanks to the other side of the field of view which serves as a control region \cite{Fomin:1994aj}. We use the multiple OFF technique \cite{HESS:2006fka} to estimate the background noise due to cosmic rays. The \textit{OFF region} or background region is defined by multiple circular regions of the same size as the \textit{ON region} or signal region which are equidistant from the pointing position, i.e.\ the centre of the cameras. As we treat our target dSph as a point-like source in the $\gamma$-ray sky, following previous CTA work \cite{CTAConsortium:2017dvg}, the size of each ON/OFF region is set to a radius of $0.1\degree$ corresponding to the average PSF of CTA. We use the Instrument Response Functions \texttt{prod3b-v2} publicly available on the CTA performance website \cite{CTAperformance} for the south site at zenith angle $z = 20\degree$, the lowest zenith angle to capture the lowest energy events, and for an observation time of 500 hours.

In this work, we focus on Sculptor, a dSph satellite of the Milky Way located in the Southern hemisphere at Galactic coordinates ($\ell=287.62\degree$, $b = -83.16\degree$) at a distance of $86 \pm 6$~kpc. The dynamics of the dSph and hence its DM density distribution is estimated based on 1365 member stars \cite{Walker:2008fc}. We make use the $J$-factor profile and its associated uncertainties provided in Ref.\ \cite{Bonnivard:2015xpq} as a function of angular radius, whose total $J$-factor reaches $\log_{10}(J/{\rm GeV}^2{\rm cm}^{-5}\rm{sr}) = 18.63^{+0.14}_{-0.08}$, assuming an Einasto DM density profile.
In the following analysis, we use the value of the $J$-factor for an angular radius of $\theta = 0.1\degree$ of Ref.\ \cite{Bonnivard:2015xpq}, $\log_{10} (J_{0.1\degree} / \text{GeV}^2 \text{cm}^{-5} \rm{sr}) = 18.3 \pm 0.3$,  corresponding to the point-like treatment of the source.

%% file: statistical_analysis.tex
% ============================================================
\section{Statistical analysis} 
\label{sec:stat_analysis}

We perform a log-likelihood ratio statistical test on the mock data in order to constrain the DM annihilation cross-section setting upper limits. We scan over the DM particle mass ranging from 30~GeV to 100~TeV divided into 100 logarithmically-spaced DM mass bins. In order to capture new features in kinematically specific regions, e.g.\ thresholds of annihilation channels or presence of a resonance, we add a selection of refined mass bins between 76 GeV and 174 GeV (see Figs.\ \ref{Fig:Gradient_BR1}, \ref{Fig:Gradient_BR2} and \ref{Fig:BR} in Sec.~\ref{Sec:SingletScalarModel} for a specific case). We assume a positive signal $\langle \sigma v \rangle > 0$, based on the method proposed in Ref.\ \cite{Cowan}. The test statistic (TS) is defined as
\begin{equation}
\rm{TS} = \left\{
\begin{array}{rcl}
&0
& \:\: \text{~~for}  \: \widehat{\langle \sigma v \rangle} > \langle \sigma v \rangle \,, \quad \\ \\
&\displaystyle{-2 \ln \frac{\mathcal{L}(\langle \sigma v \rangle, \doublehat{\boldsymbol{N}}_{\text{B}}(\langle \sigma v \rangle), \doublehat{J}(\langle \sigma v \rangle))}{\mathcal{L}(\widehat{\langle \sigma v \rangle}, \hat{\boldsymbol{N}}_{\text{B}}, \hat{J})}} & \:\: \text{~~for} \: 0 \leq \widehat{\langle \sigma v \rangle} \leq \langle \sigma v \rangle \,, \quad \\ \\
&\displaystyle{-2 \ln \frac{\mathcal{L}(\langle \sigma v \rangle, \doublehat{\boldsymbol{N}}_{\text{B}}(\langle \sigma v \rangle), \doublehat{J}(\langle \sigma v \rangle))}{\mathcal{L}(0, \doublehat{\boldsymbol{N}}_{\text{B}}(0), \doublehat{J}(0))}} & \:\: \text{~~for} \: \widehat{\langle \sigma v \rangle} < 0 \,,\quad
\end{array} \right.
\label{coef_qNS0}
\end{equation}
where $\langle \sigma v \rangle$ is the parameter of interest and ($\boldsymbol{N_\text{B}}$, $J$) are the nuisance parameters. The denominator holds the value of the annihilation cross-section $\widehat{\langle \sigma v \rangle}$, the vector of number of background events $\hat{\boldsymbol{N}}_{\text{B}}$, and $\hat{J}$ the value of the $J$-factor, that maximize unconditionally the likelihood function. The numerator contains the quantities $\doublehat{\boldsymbol{N}}_{\text{B}}(\langle \sigma v \rangle)$ and $\doublehat{J}(\langle \sigma v \rangle)$, the vector of number of background events and the $J$-factor value that maximize the likelihood function conditionally for a given annihilation cross-section $\langle \sigma v \rangle$. The upper limit on $\langle \sigma v \rangle$ for a given DM mass will be the value that responds to the criterion value of the test statistic $\rm TS$. In this work, we will derive constraints on $\langle \sigma v \rangle$ at 95\% confidence level which corresponds to a criterion value ${\rm TS} = 2.71$, in the case of a one-sided test and following previous $\gamma$-ray telescope analyses such as \cite{HESS:2021zzm, HESS:2020zwn, CTA:2020qlo, HESS:2022ygk}.

The total likelihood function $\mathcal{L}$ is the product of a Poisson likelihood $\mathcal{L}^{\mathcal P}_i$ on the events of all energy bins $i$ with a log-normal distribution $\mathcal{L}^J$ of the $J$-factor, which reads
\begin{equation}
    \mathcal{L}\big(\langle \sigma v\rangle, \boldsymbol{N_\text{B}}, J\big) ~=~ \prod_i \mathcal{L}^{\mathcal{P}}_i \big( N^i_{\text{S}}(\langle \sigma v\rangle,J), N^i_{\text{B}} \big| N^i_{\text{ON}}, N^i_{\text{OFF}},\alpha \big) \times \mathcal{L}^J(J|\bar{J},\sigma_J) \,,
\end{equation}
where $N^i_{\text{S}}$ is the number of predicted signal events for a given energy bin $i$, and $N^i_{\text{B}}$ the associated number of expected background events, with $\boldsymbol{N_\text{B}}$ the corresponding vector. The values $N^i_{\text{ON}}$ and $N^i_{\text{OFF}}$ represent the number of ON and OFF events in the energy bin $i$, respectively, and $\alpha$ is the acceptance corrected exposure ratio between both ON and OFF regions. The energy bins are logarithmically-spaced and, for the sake of sufficient statistics, they are merged with the next neighbouring one if they contain less than four ON or OFF events \cite{Feldman:1997qc}. For each energy bin $i$, $\mathcal{L}^{\mathcal{P}}_i$ is the product of two Poisson likelihood functions, corresponding to the ON and OFF regions, respectively,
\begin{equation}
\mathcal{L}^{\mathcal{P}}_i ~=~ \frac{\big(N_{\text{S}_i}(\langle \sigma v \rangle, J) + N^i_{\text{B}}\big)^{N^i_{\text{ON}}  }}{N^i_{\text{ON}}!} e^{-(N^i_{\text{S}} + N^i_{\text{B}})}
\times \frac{\big(\alpha N^i_{\text{B}}\big)^{N^i_{\text{OFF}}}}{N^i_{\text{OFF}}!} e^{-\alpha N^i_{\text{B}}} \,.
\end{equation}
Here, $N^i_{\text{S}}$ is the predicted number of signal events in the energy bin $i$ obtained through the convolution of the expected differential $\gamma$-ray flux given in Eq.\ \eqref{Eq:dPhidE} with the energy-dependent acceptance function $A_{\rm{eff}}(E_{\gamma})$, the observation time $T_{\rm{obs}}$, and the energy resolution function $R(E_{\gamma}, E'_{\gamma})$ which relates the detected energy $E'_{\gamma}$ to the true energy $E_{\gamma}$ of the events.

We then perform the integral of the convolution over the bin energy width $\Delta E_i$. The number of signal events obtained for an energy bin $i$ is computed as
\begin{equation}
 N_{\text{S}_i}\big(\langle \sigma v \rangle, J\big) ~=~  J \times \frac{1}{\xi} \frac{\langle \sigma v \rangle}{4\pi m_{\chi}^2} \int_{\Delta E_i} \int^{\infty}_0   \sum_f B_f \frac{{\rm d}N^f_{\gamma}}{{\rm d}E_{\gamma}} \: R(E_{\gamma}, E'_{\gamma}) \: A_{\rm{eff}}(E_{\gamma}) \: T_{\rm{obs}}\: {\rm d}E_{\gamma} \: {\rm d}E'_{\gamma} \,.
\end{equation}

In our analysis, we take into account the $J$-factor uncertainties with a log-normal distribution given by
\begin{equation}
    \mathcal{L}^J ~=~ \frac{1}{\ln(10) \sqrt{2\pi} \,\sigma_J \, J} \, \exp\left[-\frac{\big( \log_{10}J - \log_{10}\bar{J}\big)^2}{2 \sigma^2_J}\right] \,,
\label{L^J}
\end{equation}
where $J$ is the true value of the $J$-factor, $\bar{J}$ is the mean $J$-factor, and $\sigma_J$ is the uncertainty of $\log_{10}J$.

%% file: particle_model.tex
% ===================================================================
\section{Singlet scalar dark matter}
\label{Sec:SingletScalarModel}

In order to illustrate the impact of various annihilation channels on the limits derived from indirect dark matter detection experiments, we consider a very simple framework, where a real singlet scalar $S$ is added to the Standard Model particle content \cite{Silveira:1985rk, McDonald:1993ex}. This scalar is odd under a discrete $\mathbb{Z}_2$ symmetry and thus a viable WIMP dark matter candidate. Note that the scalar is its own antiparticle, corresponding to the case $\xi = 2$ in Eq.\ \eqref{Eq:dPhidE}. The  scalar potential is given by
\begin{equation}
    V_{\rm scalar} ~=~ \mu_H^2 |H|^2 + \lambda_H |H|^4 + \frac{1}{2} \mu_S^2 S^2 + \frac{1}{4} \lambda_S S^4 + \frac{1}{2} \lambda_{SH} S^2 |H|^2 \,.
\end{equation}
After electroweak symmetry breaking, the Higgs doublet $H$ is expressed in terms of the physical Higgs boson $h$ and the vacuum expectation value $v = \langle H \rangle \approx 246$ GeV. Moreover, minimising the potential leads to $m_h^2 = 2 \lambda_H v^2 = -2 \mu_H^2$, the Higgs mass being measured as $m_h = 125.23 \pm 0.17$ GeV \cite{PDG2022}. At tree-level, the physical mass of the singlet scalar is given by
\begin{equation}
    m_S^2 ~=~ \mu_S^2 + \frac{1}{2} \lambda_{SH} v^2 \,.
\end{equation}
The phenomenology of the model can then be fully described by only two parameters: the dark matter mass $m_S$ and the scalar coupling parameter $\lambda_{SH}$. Note that the quartic interactions $h^4$ and $S^4$ are irrelevant for dark matter phenomenology (as long as all calculations are performed at tree-level).

\begin{figure}
    \centering
    \includegraphics[width=0.9\textwidth]{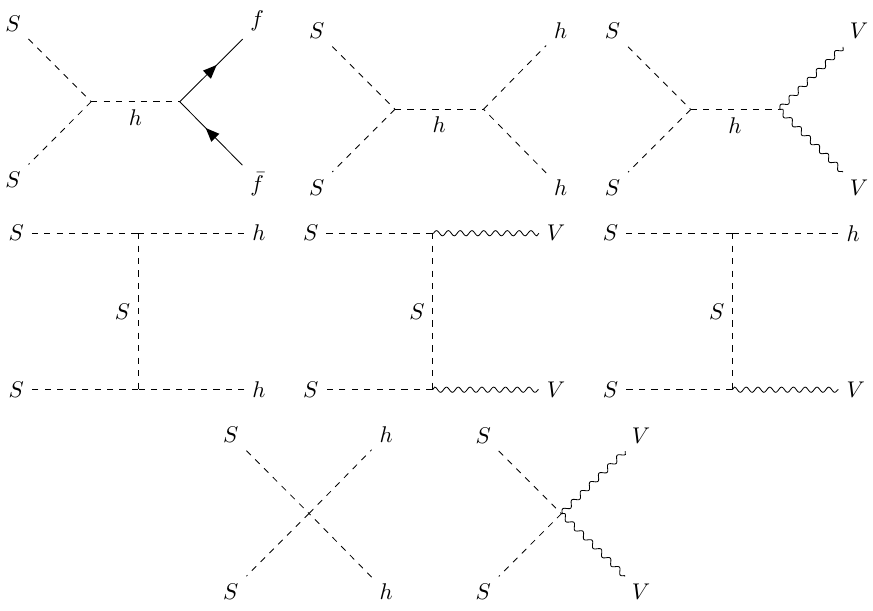}
    \caption{Tree-level Feynman diagrams for scalar pair annihilation into fermions ($f = e, \mu, \tau, u, d, c, s, t, b$), gauge ($V=W^{\pm},Z^0$) and Higgs bosons ($h$). Diagrams corresponding to $u$-channels obtained through crossing are not separately depicted. Annihilation into $\gamma\gamma$ and $gg$ final states proceeds through effective couplings to the Higgs boson $h^0$.}
    \label{Fig:SingletScalarAnn}
\end{figure}

Dark matter pair annihilation can occur into final states containing gauge and Higgs bosons, leptons, and quarks. DM annihilation into fermions proceeds solely through $s$-channel Higgs exchange, and will thus depend on the coupling parameter $\lambda_{SH}$ as well as the relevant Yukawa couplings, preferring annihilation into heavy quarks ($b$ and $t$) and $\tau$-leptons. DM annihilation into bosonic states can proceed through $s$-channel Higgs exchange, $t$- or $u$-channel singlet exchange, and through direct four-vertex interactions. Again, the parameter $\lambda_{SH}$ plays a key role in most of the contributing diagrams. Note that DM annihilations into photon ($\gamma\gamma$) or gluon ($gg$) final states involve loop-mediated diagrams and are typically included through effective couplings to the Higgs boson. All relevant Feynman diagrams are shown in Fig.\ \ref{Fig:SingletScalarAnn}.

\begin{figure}
    \centering
    \includegraphics[width=0.75\textwidth, trim={0.2cm 0.4cm 0.2cm 0.0cm},clip]{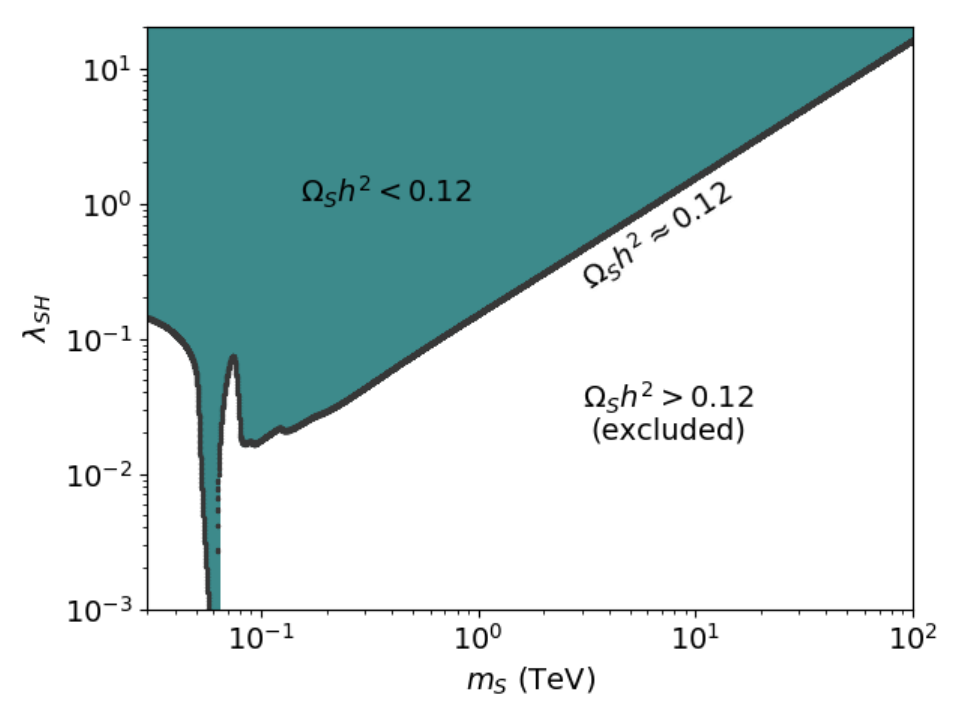}
    \caption{Dark matter relic density in the $m_S$--$\lambda_{SH}$ plane. The black line corresponds to $\Omega_S h^2 \approx 0.12$, the blue area corresponds to $\Omega_S h^2 < 0.12$. The white region is excluded because of $\Omega_S h^2 > 0.12$.}
    \label{Fig:SingletRelicDensity}
\end{figure}

\begin{figure}
    \centering
    \includegraphics[width=0.48\textwidth, trim={0.2cm 0.0cm 0.7cm 0.0cm},clip]{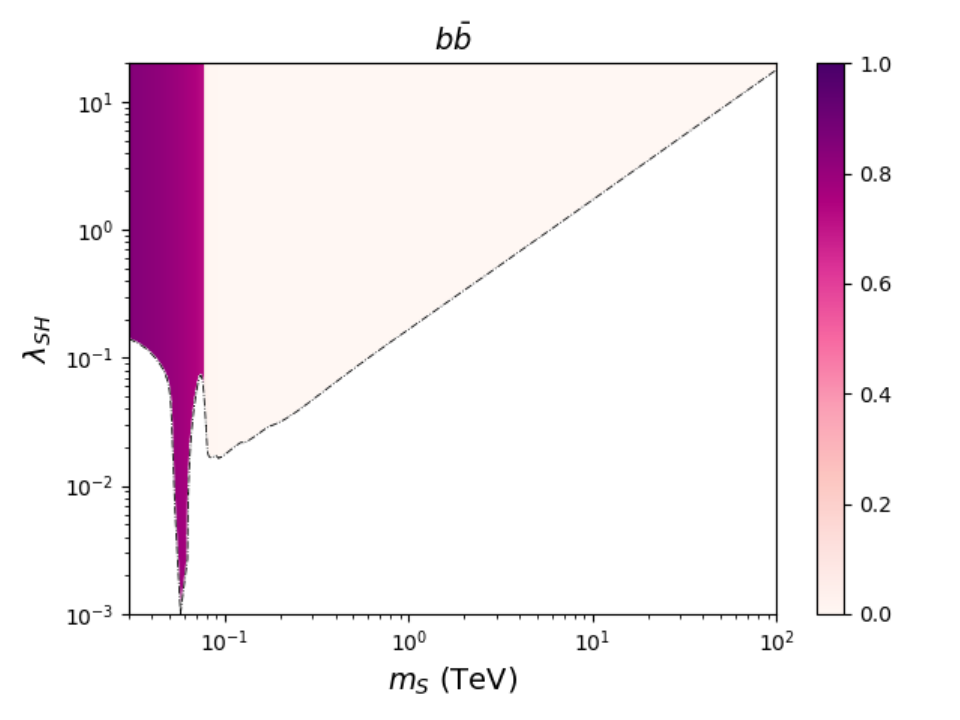}~~~
    \includegraphics[width=0.48\textwidth, trim={0.2cm 0.0cm 0.7cm 0.0cm},clip]{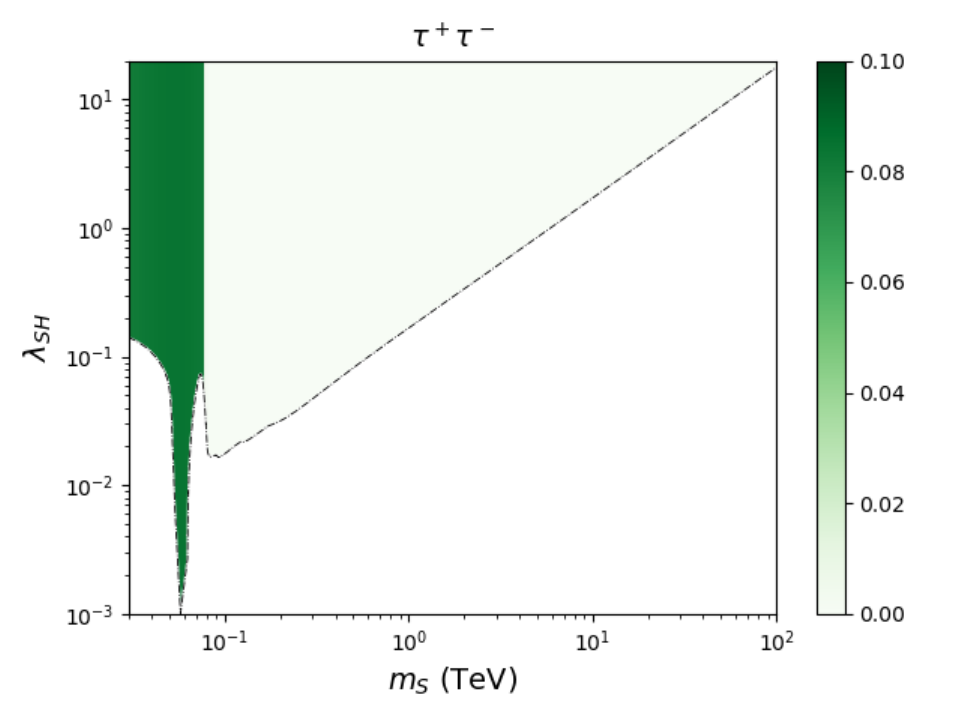}\\
    \includegraphics[width=0.48\textwidth, trim={0.2cm 0.0cm 0.7cm 0.0cm},clip]{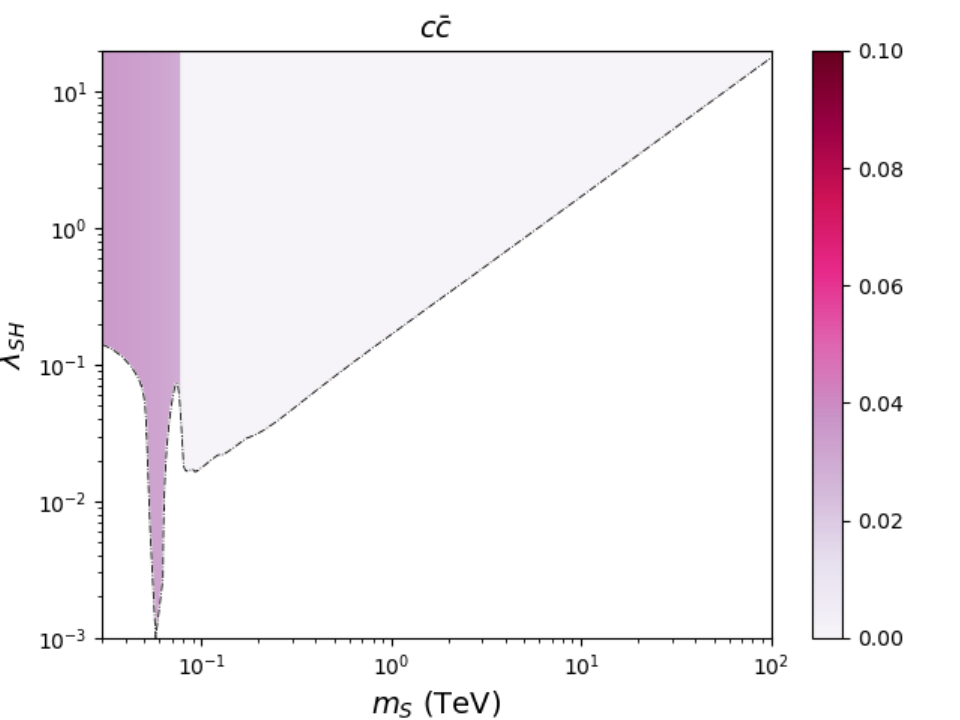}~~~
    \includegraphics[width=0.48\textwidth, trim={0.2cm 0.0cm 0.7cm 0.0cm},clip]{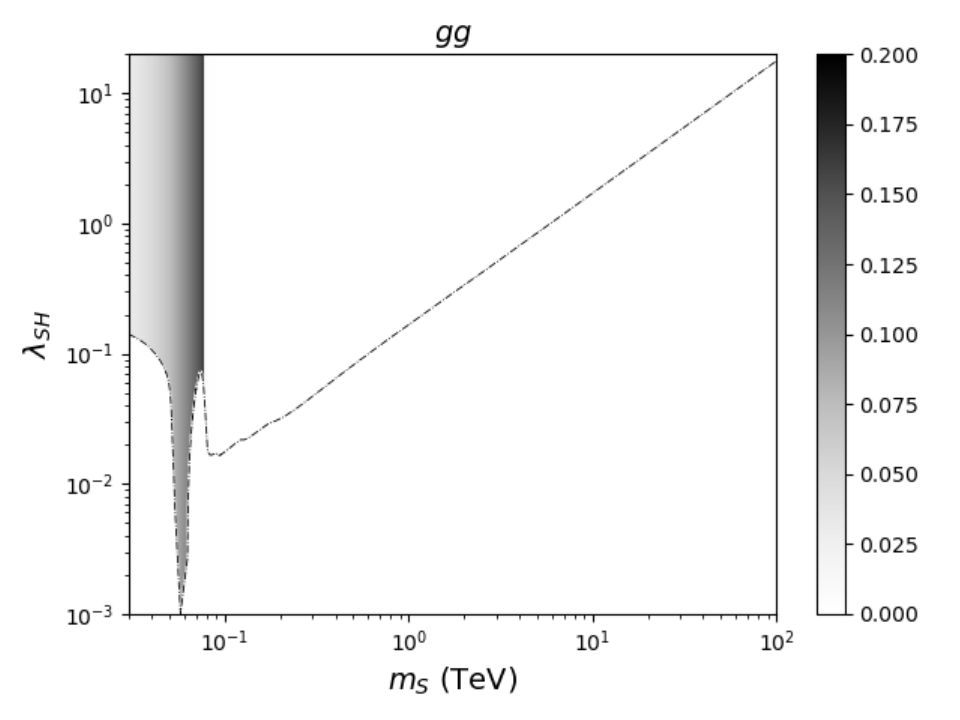}
 \caption{Gradients showing the relative contributions in terms of branching ratio of the dominant dark matter annihilation channels in the $m_S$-$\lambda_{SH}$ plane for $m_S \lesssim m_W = 80.4$ GeV. We only display parameter points leading to cosmologically viable configurations, i.e.\ $\Omega_S h^2 \leq 0.12$ (see also Fig.\ \ref{Fig:SingletRelicDensity}). The dashed lines indicate the cosmologically preferred region where $\Omega_S h^2 \approx 0.12$. Note the different scales on the colour bars. Also note that the contribution of the channels above $m_W$ is not strictly zero, but negligible for our analysis (see also Fig.\ \ref{Fig:BR}).}
    \label{Fig:Gradient_BR1}
\end{figure}

\begin{figure}
    \centering
    \includegraphics[width=0.48\textwidth, trim={0.2cm 0.0cm 0.7cm 0.0cm},clip]{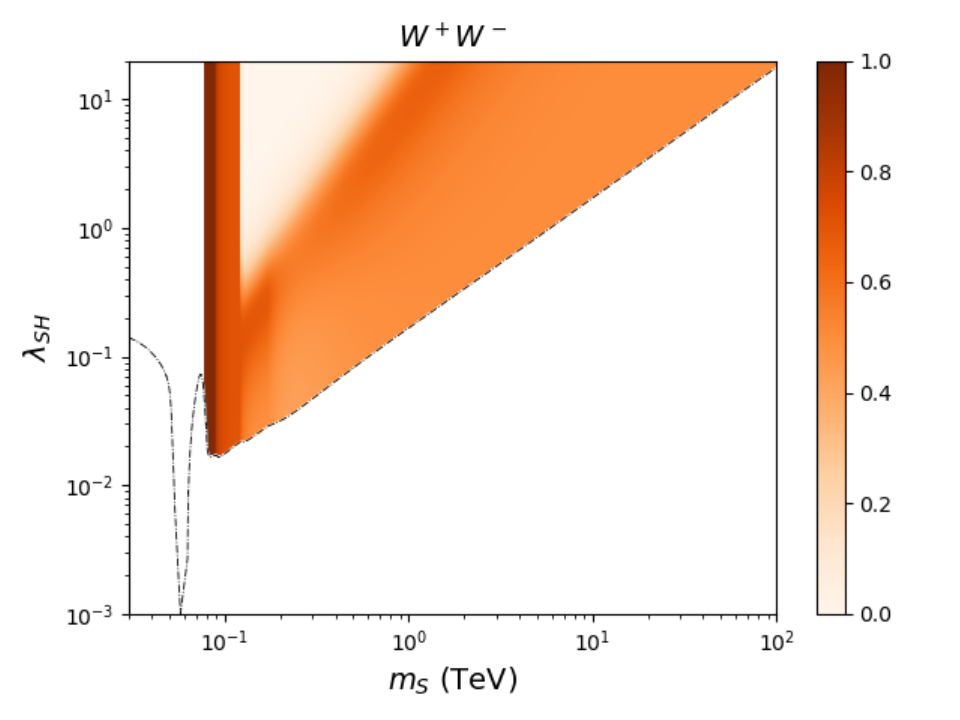}~~~
    \includegraphics[width=0.48\textwidth, trim={0.2cm 0.0cm 0.7cm 0.0cm},clip]{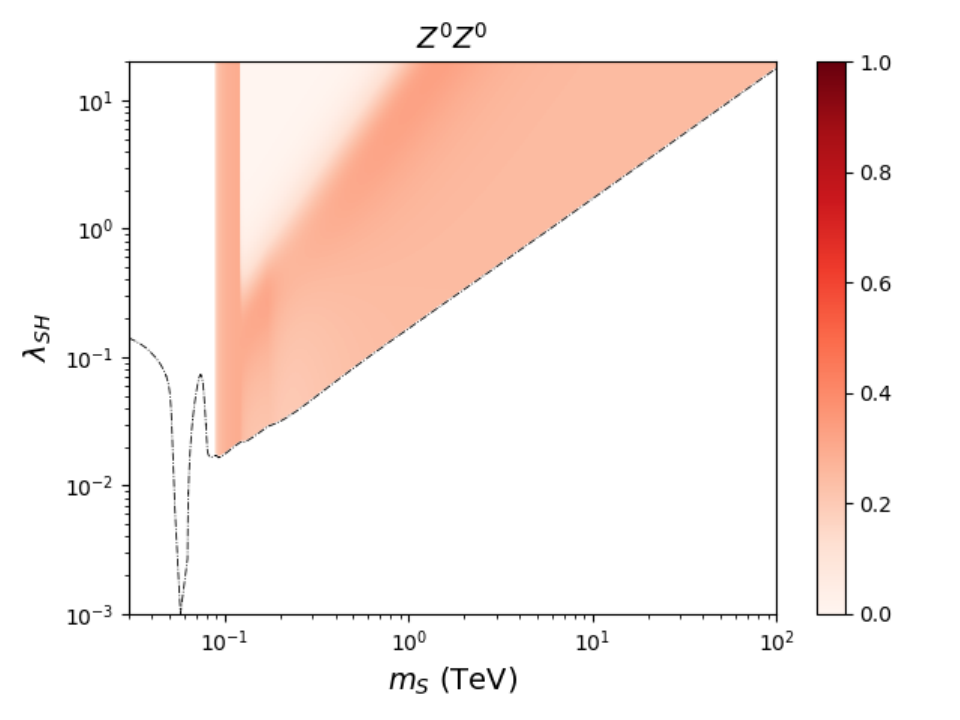}\\
    \includegraphics[width=0.48\textwidth, trim={0.2cm 0.0cm 0.7cm 0.0cm},clip]{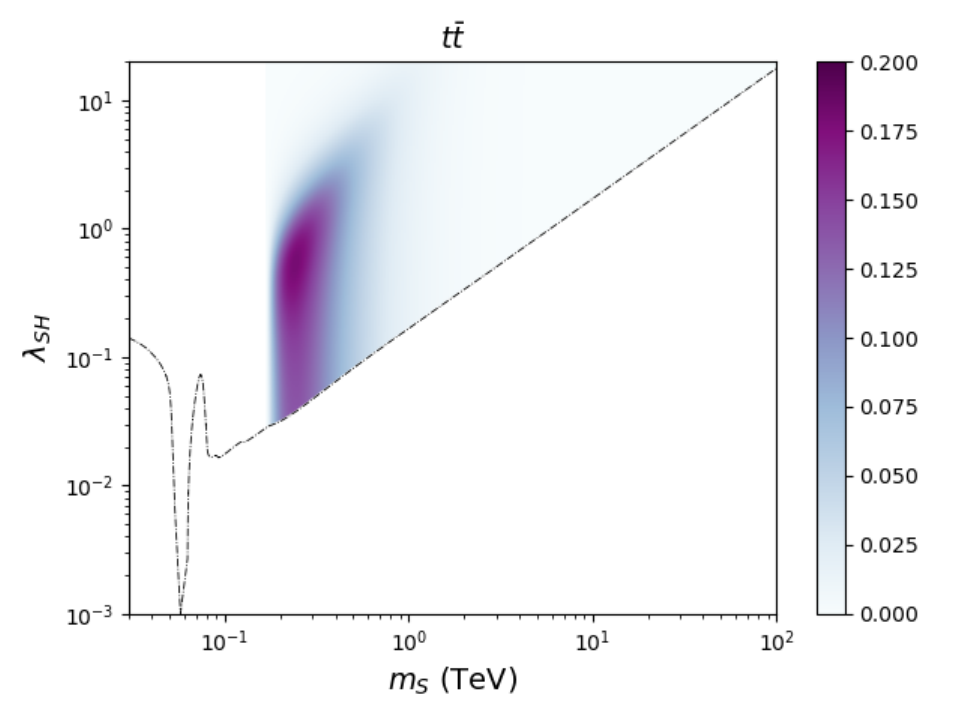}~~~
    \includegraphics[width=0.48\textwidth, trim={0.2cm 0.0cm 0.7cm 0.0cm},clip]{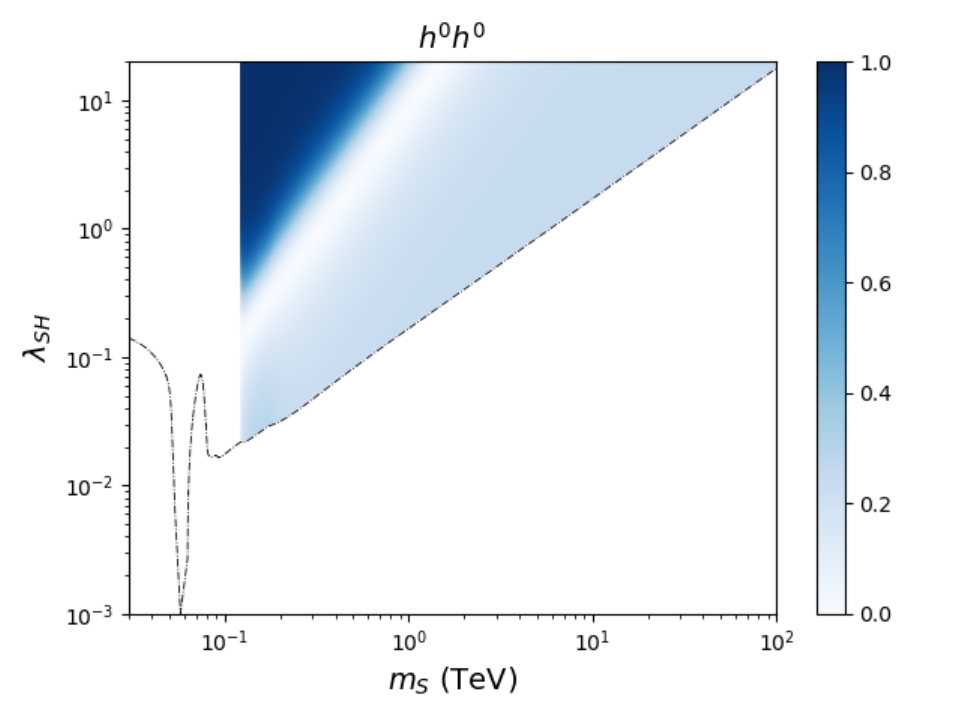}
    \caption{Same as Fig.\ \ref{Fig:Gradient_BR1} for the dominant annihilation channels for $m_S \gtrsim m_W$. Note that the channels are kinematically forbidden below their respective thresholds.}
    \label{Fig:Gradient_BR2}
\end{figure}

We make use of the package {\tt micrOMEGAs 5.2.13} \cite{MO2001, MO2004, MO2007a, MO2007b, MO2013, MO2018} which includes an implementation of the singlet scalar model to describe the DM phenomenology of the particle physics part of our study. A delicate interplay between the two key parameters $m_S$ and $\lambda_{SH}$ is needed in order to meet the stringent relic density constraint of Eq.\ \eqref{Eq:omh2Planck}. Figure \ref{Fig:SingletRelicDensity} presents the parameter space regions which are cosmologically favoured or excluded in view of the relic density constraint value, while Figs.\ \ref{Fig:Gradient_BR1} and \ref{Fig:Gradient_BR2} illustrate the most contributing DM annihilation channels in terms of branching ratios (colour bars) in the $m_S$--$\lambda_{SH}$ plane. For low dark matter masses, $m_S \lesssim 50$ GeV, the relic density constraint is met for couplings of about $\lambda_{SH} \sim {\cal O}(0.1)$. In this regime, dark matter particle annihilations occur dominantly into $b\bar{b}$ final states due to the larger Yukawa coupling, with subdominant contributions into $\tau^+ \tau^-$, $gg$ and $c\bar{c}$ as shown in Fig.\ \ref{Fig:Gradient_BR1}.

Around $m_S = m_h/2 \approx 62.5$ GeV, the Higgs-boson resonance increases the annihilation cross-section significantly. This increase has to be compensated by smaller couplings in order to maintain the singlet scalar relic density at $\Omega_S h^2 \approx 0.12$. Consequently, the value of $\lambda_{SH}$ drops as low as $10^{-4}$ in a very small mass interval around the resonance (see Fig.\ \ref{Fig:SingletRelicDensity}). After the resonance region, several kinematical thresholds are crossed at $m_S = m_W \approx 80.4$ GeV, $m_S = m_Z \approx 91.2$ GeV, $m_S = m_h \approx 125.0$ GeV, and $m_S = m_t \approx 175.3$ GeV, where the corresponding annihilation channels open up and dominate the total annihilation cross-section just above the respective kinematical threshold. Note that annihilation into $h^0h^0$ depends more strongly on the coupling $\lambda_{SH}$ leading to the observed non-uniform behaviour in the mass range between 125 GeV and approximately 1 TeV.

Finally, above $m_S \gtrsim 200$ GeV, increasing the dark matter mass requires an increase in the coupling $\lambda_{SH}$ following the relic density constraint. Here, the annihilation cross-section is dominated by the bosonic final states with $W^+ W^-$ (about 62\%), $Z^0Z^0$ (about 30\%), and $h^0h^0$ (about 8\%). While in the following we focus on indirect dark matter detection, an extensive analysis of the singlet scalar model taking into account numerous constraints has been published in Ref.\ \cite{GAMBIT:2017gge}. Let us note that although relatively large couplings $\lambda_{SH} \gtrsim 1$ may be disfavoured by arguments related to perturbativity, we include this part of the parameter space as it allows us to cover a large part of the energy range of CTA. 

Based on the various regimes described above, one can see that the assumption of one single DM annihilation channel is therefore not valid, especially around the resonance and the kinematic thresholds in the $m_S$-$\lambda_{SH}$ parameter space. 
\begin{figure}
    \centering
    \includegraphics[width=0.7\textwidth, trim={0.2cm 0.2cm 0.2cm 0.0cm},clip]{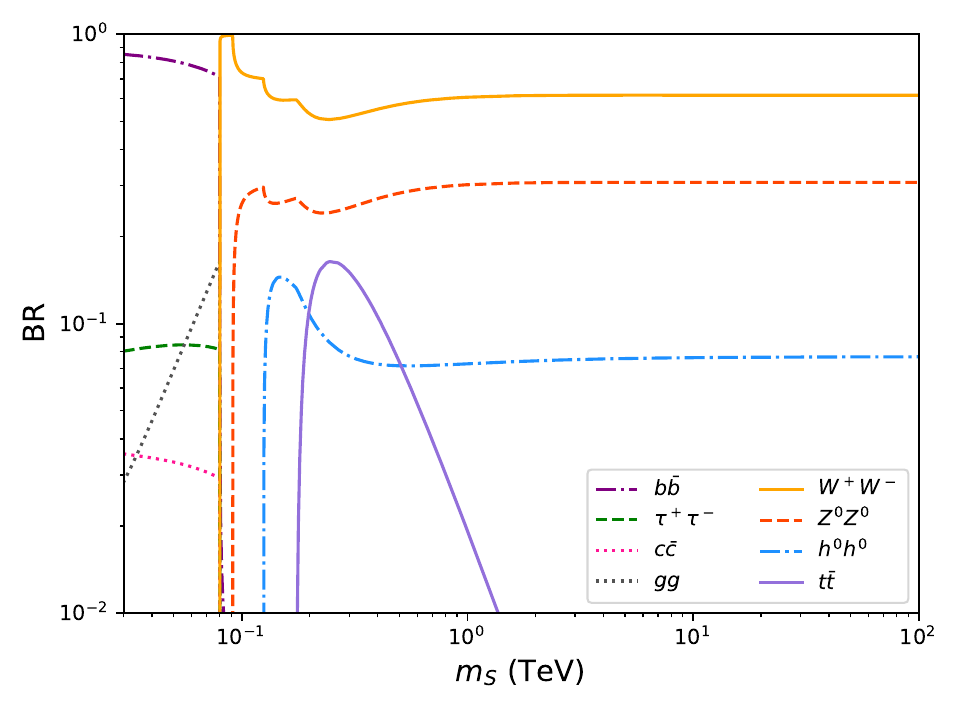}
    \caption{Branching ratios of the individual annihilation channels as a function of the dark matter mass $m_S$ when following the cosmologically preferred parameter space region where $\Omega_S h^2 \approx 0.12$ corresponding to the black band in Fig.\ \ref{Fig:SingletRelicDensity}.}
    \label{Fig:BR}
\end{figure}

In the following, we assume -- without loss of generality for our study -- that the singlet scalar accounts for the total cold dark matter present in the Universe. We are thus interested in the parameter region where $\Omega_S h^2 \approx 0.12$ according to Eq.\ \eqref{Eq:omh2Planck} manifesting as the black band in Fig.\ \ref{Fig:SingletRelicDensity}. In Fig.\ \ref{Fig:BR}, we show the different branching ratios as a function of the dark matter mass $m_S$ following precisely this parameter space region. For each mass value, the value of $\lambda_{SH}$ has been chosen such that the relic density constraint is satisfied. Again, it becomes clear that, except for the very narrow interval between $m_W$ and $m_Z$, the assumption of a single 100\% branching ratio is never satisfied. Let us finally note that, if this conclusion is found within such a minimal and simple framework, it is also expected in any extension of the Standard Model providing viable DM candidates. Dedicated interpretations within specific particle physics models are therefore at order.

%% file: results.tex
% =============================================================================
\section{Constraints on DM annihilation cross section} 
\label{sec:results}

In the absence of any significant excess found in the data obtained from the observation of, e.g., Sculptor, upper limits on the dark matter (DM) annihilation cross-section $\langle \sigma v \rangle$ can be derived as a function of the DM  mass using a log-likelihood ratio test statistic as discussed in Sec.\ \ref{sec:stat_analysis}.

In the present study, we perform the computation of predicted upper limits based on CTA mock data prepared for 500 hours of observation time. We consider the singlet scalar DM model presented in Sec.\ \ref{Sec:SingletScalarModel} assuming that the scalar field $S$ accounts for all DM present in the Universe according to Eq.\ \eqref{Eq:omh2Planck}. We take into account all relevant annihilation products -- $W^+W^-$, $Z^0Z^0$, $h^0h^0$, $gg$, $b\bar{b}$, $c\bar{c}$, $t\bar{t}$, $e^+e^-$, $\mu^+\mu^-$, $\tau^+\tau^-$, the mono-energetic channel $\gamma \gamma$, and $q\bar{q}$ including the three light quarks $u\bar{u}$, $d\bar{d}$, and $s\bar{s}$ --, all weighted by their respective branching ratio throughout the model parameter space. The differential $\gamma$-ray spectra of all annihilation channels are taken from Ref.\ \cite{Cirelli:2010xx}, obtained using {\tt PYTHIA} (version 8.135) \cite{Sjostrand:2007gs} including the final state radiative corrections. We use the mean $J$-factor value, and its uncertainty $\sigma_J$, $\log_{10}(J_{0.1\degree}/{\rm GeV}^2{\rm cm}^{-5}{\rm sr}) = 18.3 \pm 0.3$ integrated up to $\theta = 0.1\degree$ \cite{Bonnivard:2015xpq}.

\begin{figure}
    \centering{
    \includegraphics[scale=0.8, trim={0.2cm 0.2cm 0.05cm 0.0cm},clip]{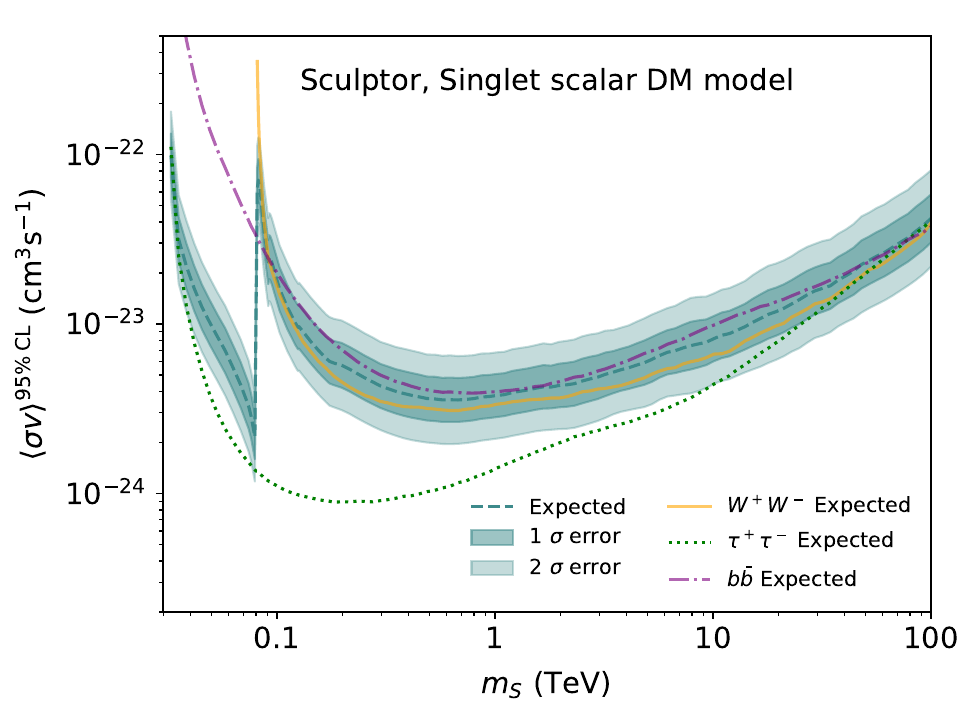}}
    \caption{Upper limits obtained at 95\% confidence level within the singlet scalar DM model taking into account the full annihilation pattern. We also show the limits obtained from three individual annihilation channels ($W^+W^-$, $\tau^+\tau^-$, and $b\bar{b}$) for comparison. The limits are presented in dependance of the dark matter mass $m_S$ following the cosmologically preferred region where $\Omega_S h^2 \approx 0.12$ corresponding to the black band in Fig.\ \ref{Fig:SingletRelicDensity}.}
    \label{Fig:UL_CTA}
\end{figure}

In Fig.\ \ref{Fig:UL_CTA}, we present the predicted upper limit and its uncertainty bands at the $1\sigma$ and $2\sigma$ confidence levels derived from a sample of 500 Poisson realizations of the background event mock data in the ON and OFF regions. The mean expected upper limit and its uncertainty bands correspond to the mean and the standard deviations at $1\sigma$ and $2\sigma$ of the $\langle \sigma v \rangle$ distribution for each DM mass. 

In the low mass regime, the limit becomes more constraining when approaching $m_S \approx m_W \approx 80$ GeV. As the DM particles get heavier, they produce more energetic SM particles which in turn generate more $\gamma$ rays. This implies a lower annihilation cross-section to compensate a higher $\gamma$-ray spectrum. We also notice an inflection point at $m_S \approx m_h/2 \approx 62.5$ GeV corresponding to the Higgs resonance. Here, the annihilation rate is increased (see Sec.\ \ref{Sec:SingletScalarModel}) such that the obtained limit decreases. 

A striking increase of the limit is then observed at $m_S \approx m_W \approx 80$ GeV, where the annihilation into the $W^+W^-$ channel opens up and dominates the total annihilation cross-section. This channel produces less $\gamma$ rays as compared to hadrons ($b\bar{b}$) (see Fig.~\ref{Fig:Cirelli_spectra_diff} in App.~\ref{sec:App_spectra_cirelli}), such that the limit increases. Our predicted upper limit on $\langle \sigma v \rangle$ reaches $3.8 \times 10^{-24}~\rm{cm}^3\,\rm{s}^{-1}$ at a DM mass of 1~TeV at 95\% confidence level. For $m_S \gtrsim 1$ TeV, the obtained limit becomes less constraining due to descreasing statistics. 

Figure\ \ref{Fig:UL_CTA} also indicates the limits obtained assuming DM particle annihilations into the individual annihilation channels $W^+W^-$, $\tau^+\tau^-$, and $b\bar{b}$, i.e.\ assuming a branching ratio of 100\% in each case. For the sake of a proper comparison, we have performed the CTA likelihood analysis on the same simulated dataset for these individual channels. We show in Fig.~\ref{Fig:Comparison_CTA} in App.~\ref{sec:App_CTA_comparison} that our predicted upper limits in the case of the individual channels are compatible with those published by the CTA collaboration \cite{CTAConsortium:2017dvg}.

While the overall shape of our limit within the singlet scalar model follows the results obtained assuming individual annihilation channels, several differences are observed:
\begin{itemize}
\item The limit obtained within the singlet scalar DM model shows to be more conservative than the one from the individual $W^+W^-$ channel. Below $m_S = m_W \approx 80.4$~GeV, no upper limit can be derived as the $W^+W^-$ channel is kinematically forbidden. Above this value, new channels open up (see Sec.\ \ref{Sec:SingletScalarModel}) and hence the total $\gamma$-ray spectrum contains additional contributions. Therefore, we observe a slight difference in favour of the $W^+W^-$ channel which produces more $\gamma$ rays than the remaining channels (see also App.\ \ref{sec:App_spectra_cirelli}). Above approximately 1 TeV, the singlet scalar DM model is dominated by annihilation into $W^+W^-$ (up to approx.\ 62\%) with subdominant contributions into $Z^0Z^0$ (approx.\ 30\%) and $h^0h^0$ (approx.\ 8\%). Here, the relative error ranges between -6\% and -22\%. Note that for masses just above the threshold, the individual $W^+W^-$ channel reaches values of relative errors beyond $100\%$.

\item In all indirect DM searches, the $\tau^+\tau^-$ channel presents the most constraining upper limits since its $\gamma$-ray production is higher than for the other channels \cite{Cirelli:2010xx}. However, in the singlet scalar DM model, the $\tau^+\tau^-$ channel is never dominant (see also Figs.\ \ref{Fig:Gradient_BR1}, \ref{Fig:Gradient_BR2} and \ref{Fig:BR} and the associated discussion in Sec.\ \ref{Sec:SingletScalarModel}). Therefore, treating $\tau^+\tau^-$ as an individual channel translates into an overestimation of the $\gamma$-ray production and consequently leads to more constraining upper limits. Below the $W$-mass threshold, the relative error varies between $12\%$ and $-65\%$. Just after the threshold, the error reaches $-98\%$ before it decreases reaching about $-3\%$ around 100 TeV.

\item Regarding the hadronic channel $b\bar{b}$, the limit obtained within the singlet scalar model is generally more stringent than the one obtained considering this channel alone. For $m_S \lesssim m_W$, we observe an important difference between the two limits of slightly more than one order of magnitude. This is explained by the albeit subdominant presence of the $\tau^+\tau^-$ channel, which yields a larger amount of $\gamma$ rays. Although $\tau^+\tau^-$, $c\bar{c}$, and $gg$ account for maximally about 25\% of the total annihilation cross-section, their contribution decreases the obtained limits in a significant way. For $m_S \gtrsim m_W$, the $b\bar{b}$ channel is suppressed in the singlet scalar model. As for the $\tau^+\tau^-$ channel discussed above, considering this channel alone leads to an inaccurate estimation of the upper limit. Here, deriving the limit based on $b\bar{b}$ alone yields a less constraining upper limit due to its softer $\gamma$-ray spectrum. In this case, the relative error below the $W$ mass is of the order of $\mathcal{O}(1000\%)$ due to the important discrepancies between the two limits. The error then drops to $-56\%$ at the $W$ mass, then remains in the range between $-10\%$ and $+38\%$ after the $W$ threshold.
\end{itemize}

\begin{figure}
    \centering
    \includegraphics[width=0.7\textwidth, trim={0.2cm 0.2cm 0.2cm 0.0cm},clip]{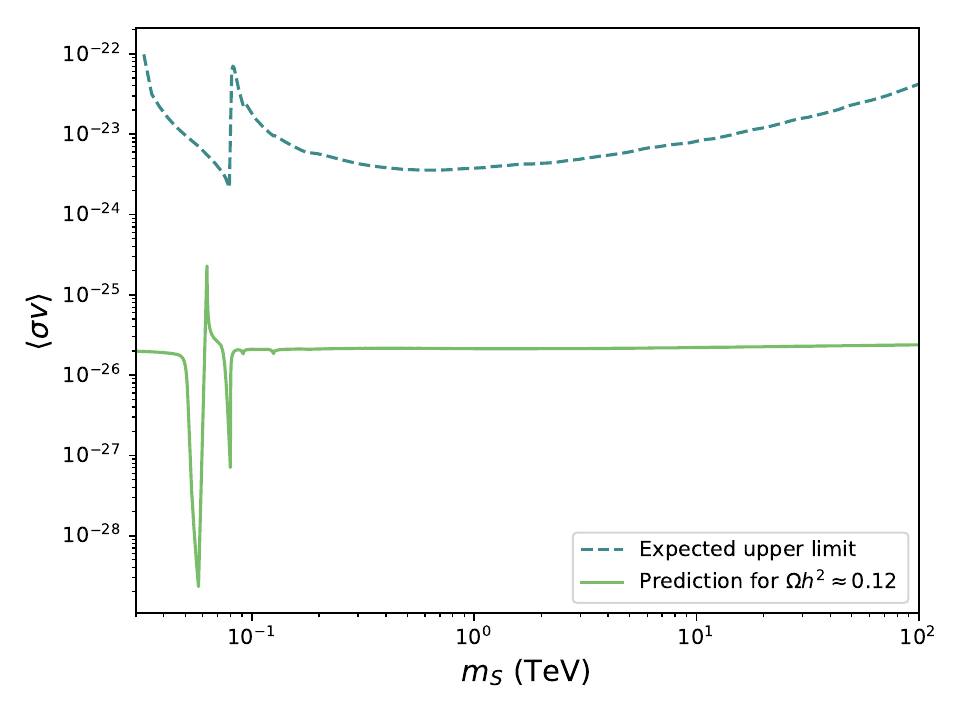}
    \caption{Dark matter annihilation cross-section (solid green line) in the singlet scalar model compared to the expected limit (dashed blue line) presented in Fig.\ \ref{Fig:UL_CTA} as a function of the dark matter mass $m_S$ following the parameter space region where $\Omega_S h^2 \approx 0.12$ corresponding to the black band in Fig.\ \ref{Fig:SingletRelicDensity}.}
    \label{Fig:XSec}
\end{figure}

Let us note that combining the individual limits (obtained from the individual annihilation channels assuming a 100\% branching ratio) by simply reweighting them with the corresponding branching ratios from the particle physics models does not lead to an accurate estimation of the complete limit on the annihilation cross-section for all DM masses. While such an approximation may be reasonable in the case where the contributing channels feature similar $\gamma$-ray spectra (e.g.\ in the singlet scalar model for $m_S \gtrsim 1$ TeV), it will not be valid in the case of rather different spectra (e.g.\ singlet scalar model for $m_S \lesssim m_W$). 

We finally show in Fig.\ \ref{Fig:XSec} the comparison of the obtained limit, taking into account the full model information, and the predicted total annihilation cross-section within the singlet scalar model. Although the model would not be excluded by the presented limit, the graph illustrates again to which extend the resonance and the kinematical thresholds affect both the total annihilation cross-section and the expected exclusion limit. In a situation where the two curves are generally closer one to the other, the observed fluctuations may easily lead to an exclusion in the corresponding mass range.

%% file: conclusion.tex
% ===================================================================
\section{Conclusions} \label{sec:conlusions}

Current limits on indirect dark matter detection cross-sections are mainly derived based on the assumption of one single annihilation channel and without considering specific particle physics models. We have first demonstrated that this assumption is not valid within a given framework providing a viable candidate for WIMP dark matter. In the singlet scalar dark matter model, the typical channels $b\bar{b}$ and $W^+W^-$ dominate the annihilation cross-section in only a restricted part of the viable parameter space, while, e.g., annihilations into $\tau^+\tau^-$ and $t\bar{t}$ remain subdominant in this model. Second, we have shown that taking into account the full annihilation pattern of the WIMP particle can have a significant impact on the derived limits of the dwarf spheroidal galaxy Sculptor which can shift by more than an order of magnitude. Depending on the exact situation, the obtained limits may be more or less constraining than those from individual channels only. Based on these results, one can see that it is necessary to take into account all annihilation channels producing different $\gamma$-ray spectral shapes to capture additional features in the DM annihilation cross-section upper limits. This conclusion can also be drawn, e.g., from Refs.\ \cite{GAMBIT:2018eea,Baldes:2020hwx} which focus on the reinterpretation of the indirect detection results in the context of specific models including specific energy spectra.

Our analysis has been performed using CTA mock data of Sculptor. A similar impact can naturally be expected for other $\gamma$-ray sources as well as for other $\gamma$-ray observatories. The numerical setup that has been elaborated for this study, namely combining the particle physics and the astrophysics aspects, could be used on the future observations of the CTA dark matter programme or on the data of any $\gamma$-ray experiment. 

Let us finally point out that our demonstration has been carried out in a very simple particle physics model, where the Standard Model is extended by only a singlet scalar, which is the WIMP dark matter candidate. Even in this setup the assumption of a single annihilation channel basically never holds. Consequently, it generally cannot be expected to hold in more complex extensions of the Standard Model involving a richer field content or even several possible dark matter candidates such as, e.g., supersymmetric models \cite{Ellis:1983ew, Martin:1997ns}, the inert doublet model \cite{Deshpande:1977rw, LopezHonorez:2006gr, Goudelis:2013uca}, or scotogenic models \cite{Restrepo:2013aga, Esch:2018ccs, Sarazin:2021nwo}. We suggest to derive upper limits on the dark matter annihilation cross-section within concrete particle physics frameworks rather than considering generic individual annihilation channels.

%% file: App_spectra_cirelli.tex
\newpage
\section{Spectrum comparison of the contributing channels} 
\label{sec:App_spectra_cirelli}

\begin{figure}[h!]
    \centering
    \includegraphics[width=0.48\textwidth, trim={0.2cm 0.0cm 1.6cm 0.0cm},clip]{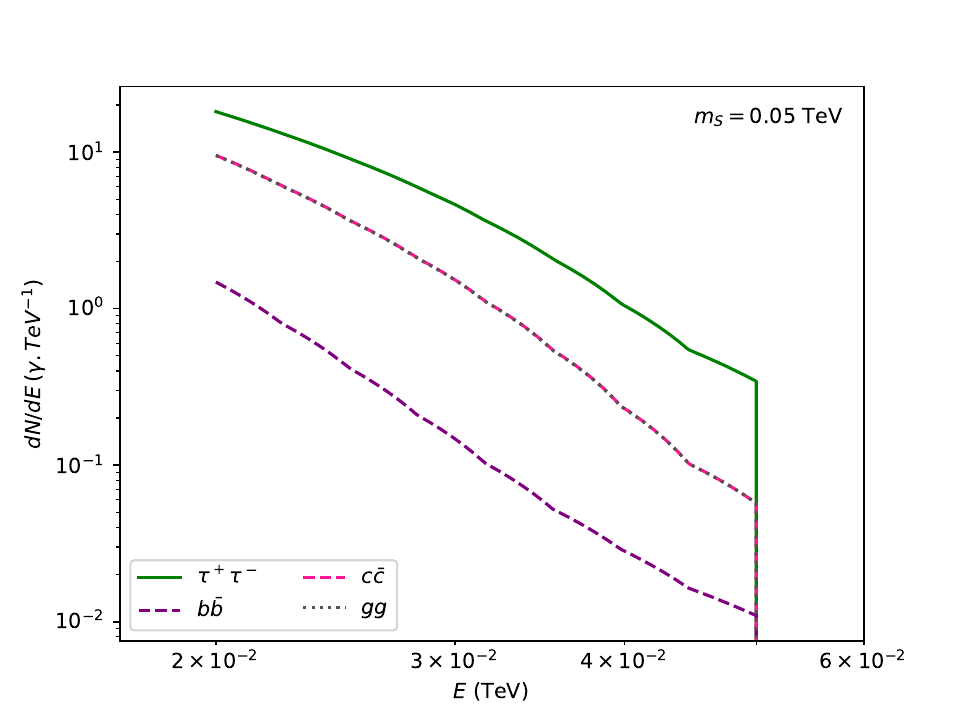}~~~~
    \includegraphics[width=0.48\textwidth, trim={0.2cm 0.0cm 1.6cm 0.0cm},clip]{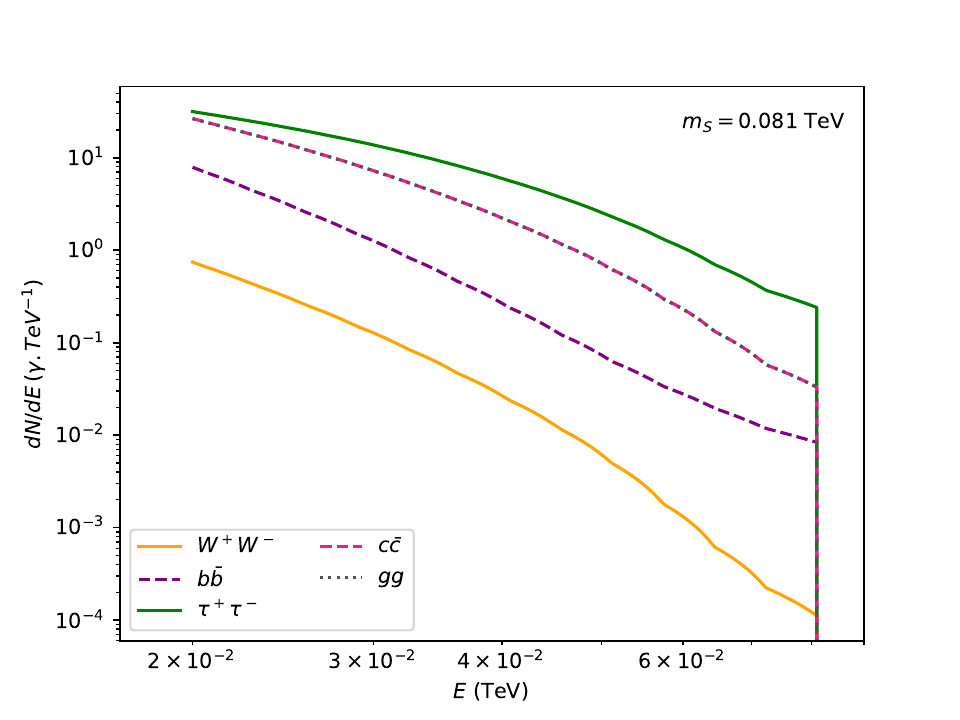}
    \includegraphics[width=0.48\textwidth, trim={0.2cm 0.0cm 1.6cm 0.0cm},clip]{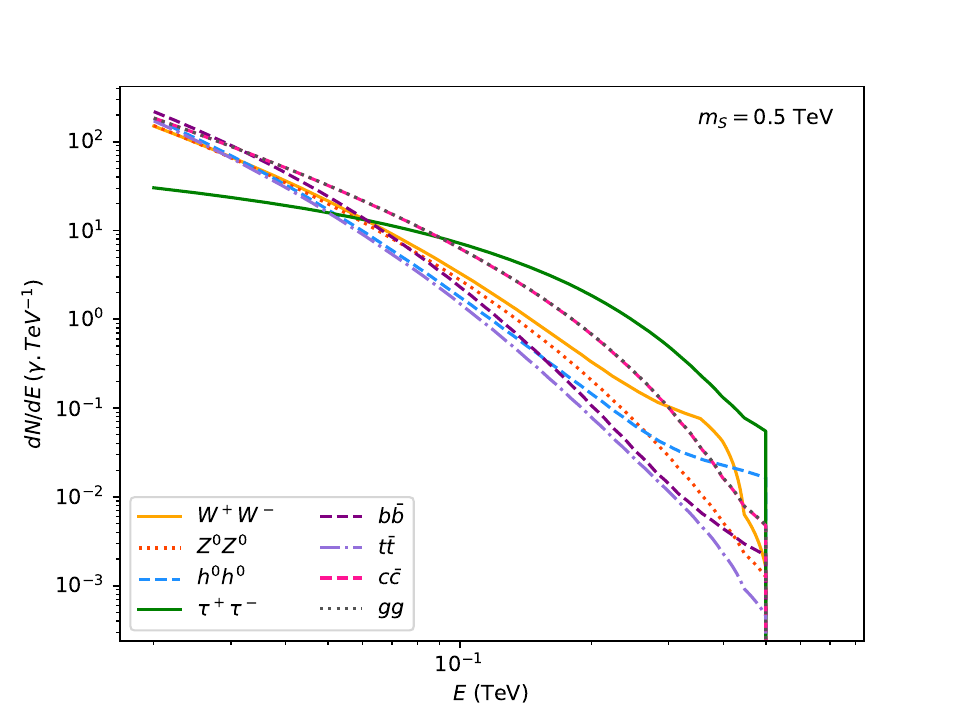}~~~~
    \includegraphics[width=0.48\textwidth, trim={0.2cm 0.0cm 1.6cm 0.0cm},clip]{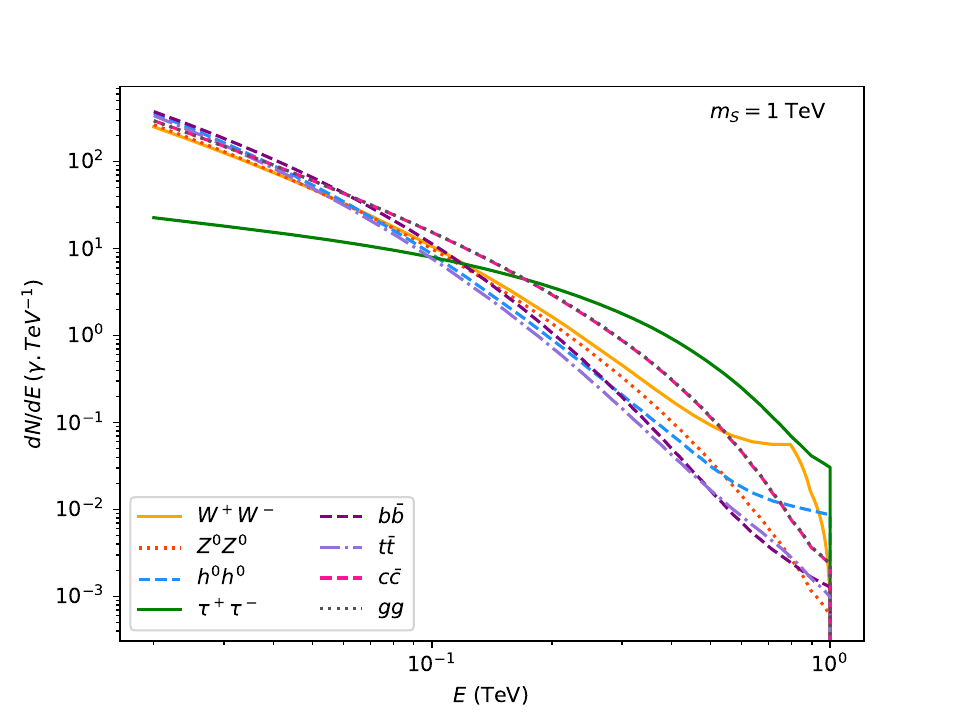}
    \includegraphics[width=0.48\textwidth, trim={0.2cm 0.0cm 1.6cm 0.0cm},clip]{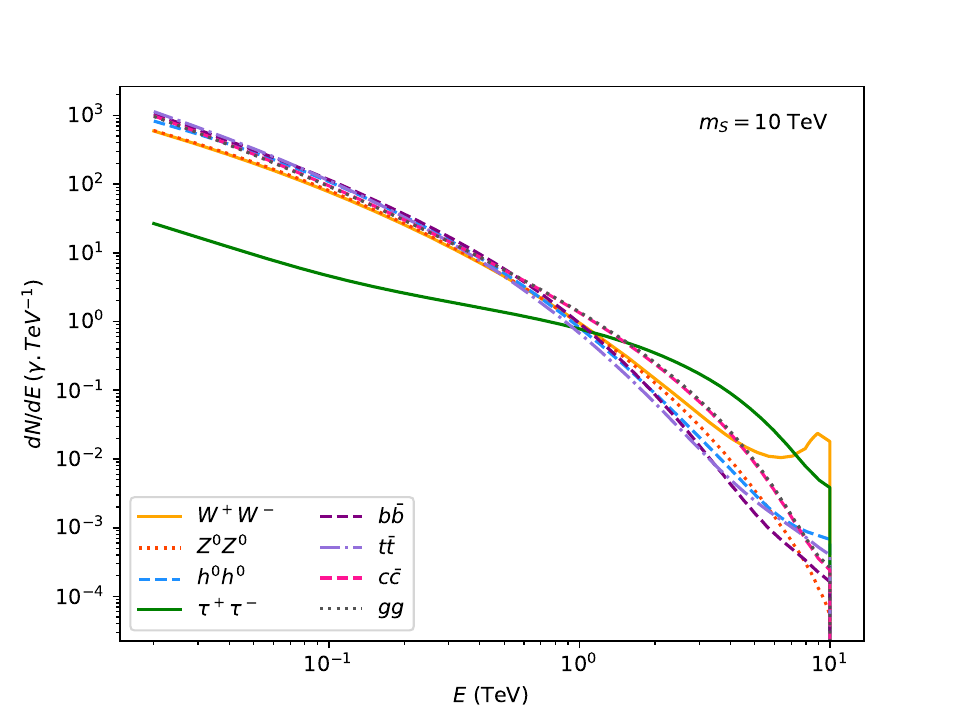}~~~~
    \includegraphics[width=0.48\textwidth, trim={0.2cm 0.0cm 1.6cm 0.0cm},clip]{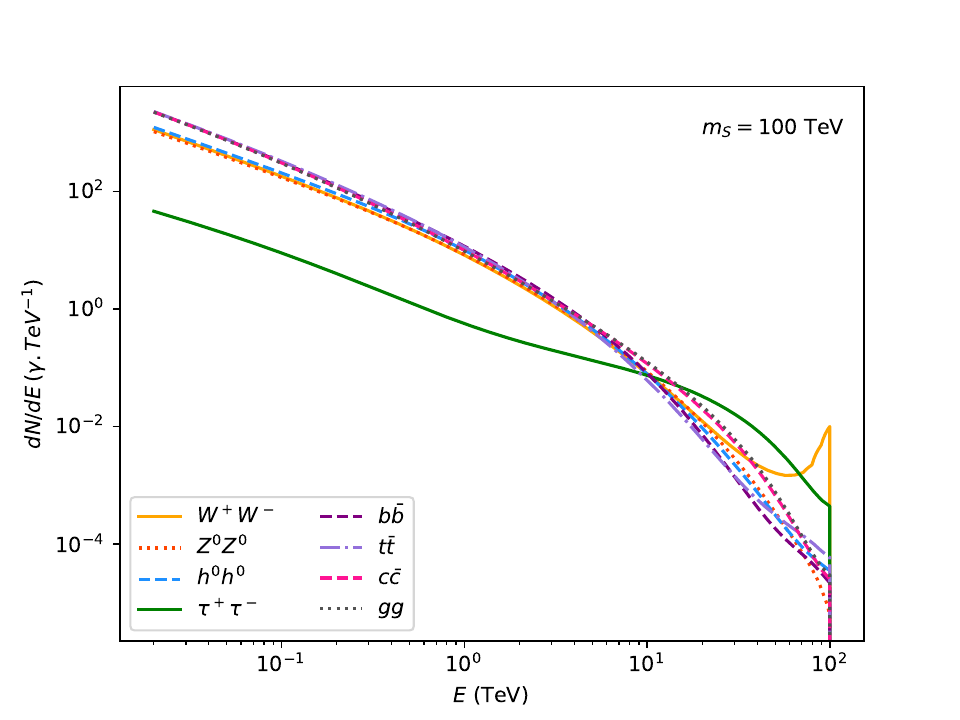}
    \caption{Differential $\gamma$-ray energy spectra of the most contributing dark matter annihilation channels at $m_S = 0.05$, $0.081$, $0.5$, $1$, $10$, and $100$~TeV.}
    \label{Fig:Cirelli_spectra_diff}
\end{figure}

%% file: App_CTA_comparison.tex
\section{Comparison with CTA results} 
\label{sec:App_CTA_comparison}

\begin{figure}[h!]
    \centering
    \includegraphics[width=0.48\textwidth]{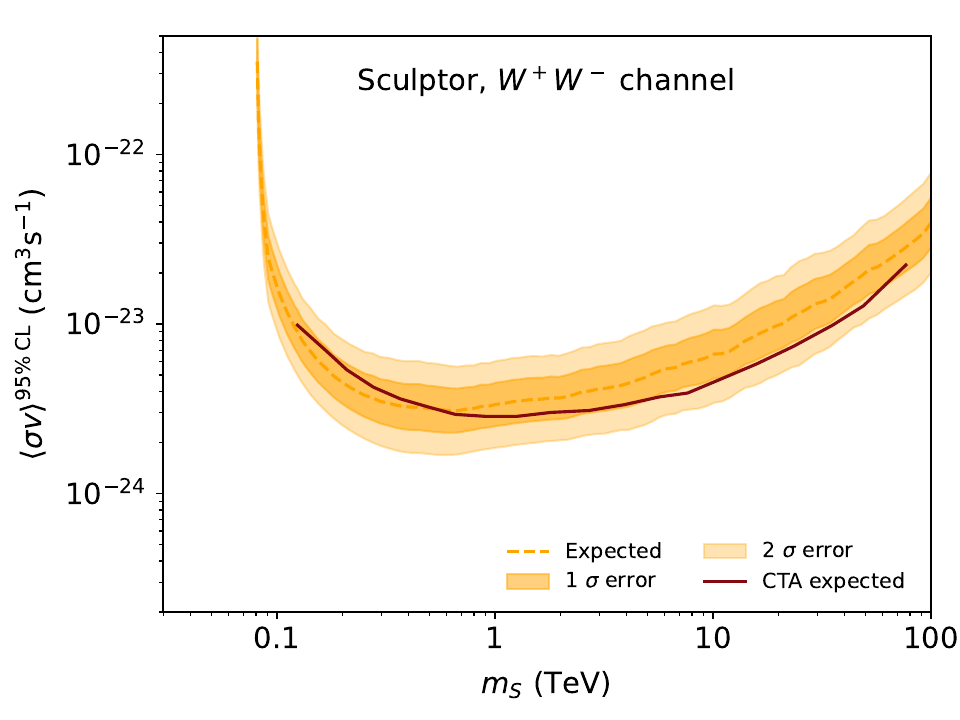}~~~~
    \includegraphics[width=0.48\textwidth]{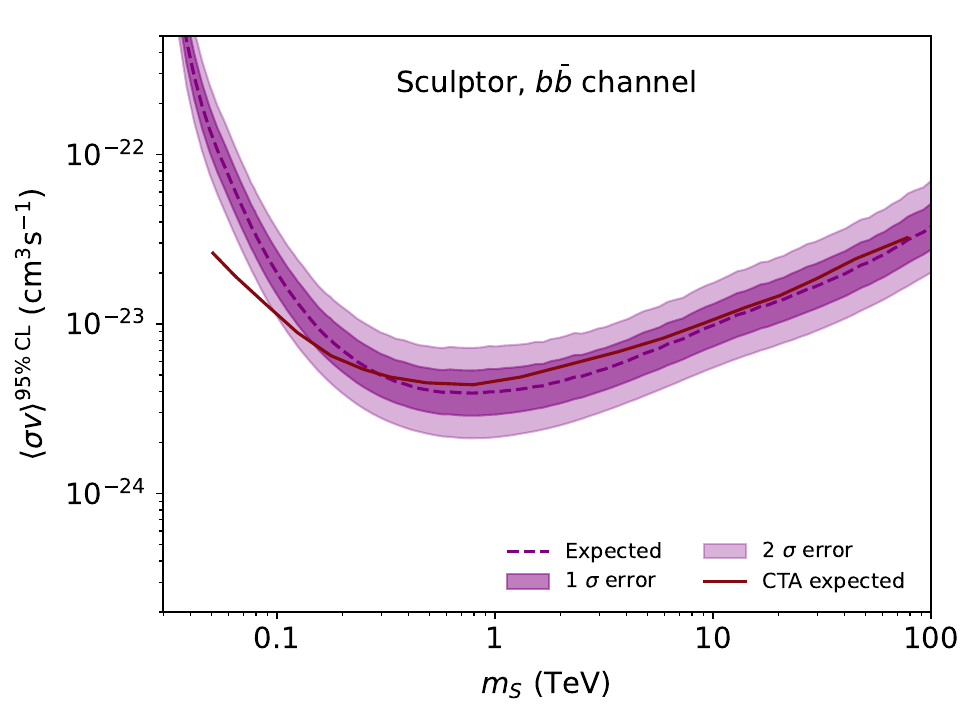}
    \includegraphics[width=0.48\textwidth]{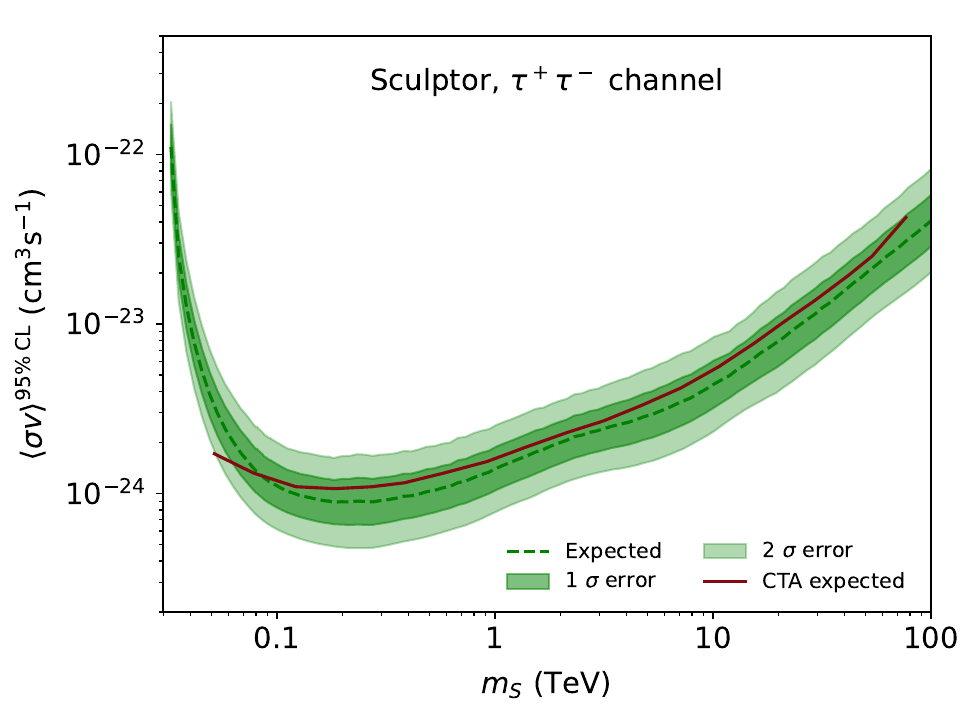}
    \caption{Comparison of our expected upper limits on the DM annihilation cross-section to those predicted by the CTA collaboration \cite{CTAConsortium:2017dvg} for the individual annihilation channels $W^+W^-$, $b\bar{b}$, and $\tau^+ \tau^-$ for 500h of observation. Our results are in agreement with the results of CTA (solid red lines) within the $1\sigma$ and $2\sigma$ uncertainty bands.}
    \label{Fig:Comparison_CTA}
\end{figure}